\documentclass[sigconf, screen, authorversion]{acmart}

\RequirePackage{rotating, adjustbox}
\AtBeginDocument{%
  }

\setcopyright{acmcopyright}
\copyrightyear{2024}
\acmYear{2024}
\acmDOI{XXXXXXX.XXXXXXX}

\acmConference[ETRA '24]{ACM Symposium on Eye Tracking Research \& Applications}{June 04--07,
  2024}{Glasgow, UK}
\acmSubmissionID{1075}

\begin{document}

\title[On Task and in Sync]{On Task and in Sync: Examining the Relationship between Gaze Synchrony and Self-Reported Attention During Video Lecture Learning}

\author{Babette Bühler}
%\authornote{}

\orcid{0000-0003-1679-4979}

\affiliation{%
  \institution{Hector Research Institute of Education Sciences, University of Tübingen}
  \streetaddress{Europastraße 6}
  \city{Tübingen}
  \country{Germany}
  \postcode{72072}
}
\email{babette.buehler@uni-tuebingen.de}

\author{Efe Bozkir}
\affiliation{%
  \institution{University of Tübingen \& Technical University of Munich}
  \country{Germany}}
\email{efe.bozkir@tum.de}
\orcid{0000-0002-4594-4318}

\author{Hannah Deininger}
\affiliation{%
\institution{Hector Research Institute of Education Sciences, University of Tübingen}
  \city{Tübingen}
  \country{Germany}
}
\email{hannah.deininger@uni-tuebingen.de}
\orcid{0000-0002-8754-6123}

\author{Peter Gerjets}
\affiliation{%
 \institution{Leibniz-Institut für Wissensmedien}
 \streetaddress{Schleichstraße 6}
 \city{Tübingen}
 \postcode{72076}
 \country{Germany}}
 \email{p.gerjets@iwm-tuebingen.de}
\orcid{0000-0003-1358-6779}

\author{Ulrich Trautwein}
\affiliation{%
\institution{Hector Research Institute of Education Sciences, University of Tübingen}
  \city{Tübingen}
  \country{Germany}}
\email{ulrich.trautwein@uni-tuebingen.de}
\orcid{0000-0003-0647-0057}

\author{Enkelejda Kasneci}
\affiliation{%
  \institution{Technical University of Munich}
  \streetaddress{Arcisstraße 21}
  \city{Munich}
  \country{Germany}
  \postcode{80333}}
\email{enkelejda.kasneci@tum.de}
\orcid{0000-0003-3146-4484}

\renewcommand{\shortauthors}{Bühler et al.}

\begin{abstract}
  Successful learning depends on learners’ ability to sustain attention, which is particularly challenging in online education due to limited teacher interaction. A potential indicator for attention is gaze synchrony, demonstrating predictive power for learning achievements in video-based learning in controlled experiments focusing on manipulating attention. This study (N=84) examines the relationship between gaze synchronization and self-reported attention of learners, using experience sampling, during realistic online video learning. Gaze synchrony was assessed through Kullback-Leibler Divergence of gaze density maps and MultiMatch algorithm scanpath comparisons. Results indicated significantly higher gaze synchronization in attentive participants for both measures and self-reported attention significantly predicted post-test scores. In contrast, synchrony measures did not correlate with learning outcomes. While supporting the hypothesis that attentive learners exhibit similar eye movements, the direct use of synchrony as an attention indicator poses challenges, requiring further research on the interplay of attention, gaze synchrony, and video content type.

\end{abstract}






\maketitle

\section{Introduction}

The increasing transition from traditional classroom settings to digital learning environments changes student-teacher interactions, which are crucial for learning. Without a physically present instructor to tailor content and provide support, learners need to self-regulate to sustain focus on educational tasks \cite{schacter2015}. This presents a significantly greater challenge for students in a video lecture than in a traditional face-to-face lecture \cite{wammes2017}. Although online teaching can take place in real-time, it proves more difficult for lecturers to monitor and manage learners' attention, causing a desire for teachers to receive feedback \cite{sun2019}. This has led to extensive research on assessing attention in online learning \cite{schacter2015, wammes2019, wisiecka2022}. 
Eye gaze is one of the most crucial sensing modalities for measuring human attention. As such it has been extensively studied also in the educational context~\cite{goldberg2021attentive, sumer2018teachers, castner2018scanpath, gao2021digital, bozkir2021exploiting}. Recent studies showed that visualizations of gaze data enable instructors to estimate the level of attention of learners \cite{sauter2023} and their comprehension of learning content \cite{Kok.2023}. In a pioneering study, \cite{Madsen.2021} hypothesized that attentive learners follow instructional videos similarly with their eyes. Measuring synchrony with inter-subject correlation (ISC) of eye gaze within experimental groups of attentive and distracted students, they showed that gaze synchrony levels were predictive of test scores. Additionally, they expanded their approach to webcam-based eye-tracking data, suggesting the potential for real-time assessment of attention and adaption of learning content. In a later study, \cite{Sauter.2022} could not reproduce these findings with unreliably sampled data, highlighting the challenges of webcam-based eye-tracking for educational contexts. In another study, \cite{Liu.2023} only found weak correlations between gaze synchrony and test scores and decreased synchrony when using eye gaze models, whereas confirming previous findings that eye gaze models foster learning \cite{jarodzka2013}.

A major limitation of previous research is that participants' attention level was experimentally controlled by introducing a secondary distraction task in the inattentive condition~\cite{Madsen.2021, Liu.2023}. It remains uncertain whether these experimental manipulations accurately reflect naturally occurring distractions \cite{Liu.2023}. The underlying mechanisms of inattention during video learning are multifaceted and range from overt distraction by the environment, for example, an incoming email \cite{wammes2019}, to hidden cognitive processes such as mind wandering, i.e., the engagement in task-unrelated thought, for example, daydreaming \cite{lindquist2011}. 
In realistic learning scenarios, the level of attention can be assumed to exhibit more gradation. Distraction and cognitive disengagement are dynamic processes that evolve and fluctuate over time \cite{farley2013}. Intruding thoughts may be less persistently inhibitory, potentially less evident in eye movements than the artificially induced counting task, yet still detrimental to learning. Further, only the very time-sensitive frame-wise ISC has been employed as a synchrony measure, suitable for very dynamic stimuli, which are not necessarily given in the educational context. A widespread video lecture setup presents an instructor and slides.

In this paper, we investigate the hidden link between gaze synchrony, attention, and learning outcomes. We examine the relationship between gaze synchrony and self-reported attention in a realistic learning scenario, specifically during learning with a pre-recorded Zoom video lecture. Gaze synchrony is assessed with two different measures: the Kullback-Leibler divergence between gaze density maps and the MultiMatch scanpath comparison method. Further, we examine the suitability of gaze synchrony as an indicator of attention by investigating the relationship between gaze synchrony, self-reported attention, and learning outcomes in the form of post-test scores. Our primary contributions include investigating naturally occurring self-reported (in)attention and its association with gaze synchrony, employing novel measures for gaze synchronization in video learning, and examining whether gaze synchrony predicts learning outcomes in this setting.

% so far: gaze synchrony predicts test score (assumed to indicate attention, only showed with experimental induced distractions)

% ! Missing link ->  does gaze synchrony signify "naturalistic" attention in the learning context
% •	Synchrony is indicator for attention ( Explain link between gaze synchrony and performance)
% •	Realistic scenario using 60 minute long online lecture
% •	Does synchrony predict attention/ being On task as a mediator for test scores?

% gaze driven by dynamic of video: slide format rather static videos, still works?

%Since our video stimulus consists entirely of the rather static slides of a PowerPoint presentation and the only dynamic components guiding the viewers gaze are movements of a cursor functioning as a mouse and the webcam video of the lecturer.

\section{Related Work}
\subsection{Synchrony during Learning}

In their study, \cite{Madsen.2021} proposed that learners synchronize their gaze during cognitive processing of lecture videos, indicating attentiveness. To investigate this, participants in two experimental conditions, attentive and distracted, viewed short informal instructional videos. In the distracted condition, participants were instructed to do the serial subtraction task (counting back in their minds from a randomly chosen prime number in decrements of 7) while watching the video. The attentive condition did not get any extra instructions. The synchronicity of eye movements was significantly higher in the attentive than in the distracted condition. They found a significant correlation between the level of gaze synchronicity and test scores, and the results were robust across factual and comprehension questions, as well as different video styles (animation vs. drawing figures). 

In replicating the prior study, including various lecture video styles and using webcam-based eye tracking, without extensive exclusion of low sampling rates, \cite{Sauter.2022} did not observe a predictive correlation between gaze ISC and test scores in a comprehensive quiz. This underscores the constraints of current webcam-based eye-tracking methods and their limitations in employing a synchrony measure in real-time remote settings. Additional studies with high-end eye trackers confirmed the correlation between experimentally induced distraction and ISC \cite{Liu.2023}. In a second experiment, however, \cite{Liu.2023} showed that the use of gaze modeling decreased the level of gaze synchrony compared to a normal viewing condition while still showing higher post-test results. A very weak correlation between total gaze synchrony and test scores was found.

These studies strongly suggest the significance of gaze synchronization during video learning, correlating with learning success, and suggest an interrelation to attention. Previous research investigating gaze patterns in live online lectures reinforces these observations, highlighting a positive correlation between students' focal attention and their ability to retain lecture content \cite{wisiecka2022}. However, it remains unclear if gaze (dis-)synchrony can indeed be used as an indicator for naturally occurring (in-)attention and employed for attention detection in a real-world learning scenario. 

\subsection{Measuring Gaze Synchrony}

The only measure employed to date to assess gaze synchrony during learning is the ISC of vertical and horizontal gaze positions and pupil diameter per video frame \cite{Madsen.2021, sauter2023, Liu.2023}. It is computed within the experimentally induced attentive and distracted groups. In the webcam-based eye tracking setting, pre-computed median values and eye movement velocity instead of pupil diameter were used for correlation computation~\cite{Madsen.2021,sauter2023}. The ISC is a very time-sensitive measure that assumes synchronization with minimal spatial-temporal distances, which appears to be suitable for a very dynamic video stimulus, which more strongly drives eye movements \cite{Dorr.2010}. This raises the question of the applicability of ISC for more static settings in the educational domain. A frequently selected format, particularly in live teaching settings, is the presentation of lecture slides. This represents a relatively static stimulus, in which the gaze is mainly guided by the lecturer's verbal description of the content and potential pointers like cursors \cite{sauter2023}.  

Various other measures have been proposed in different contexts to capture the synchronicity of eye movements in dynamic scenes. This includes clustering-based approaches, measuring the percentage of gaze falling into the main cluster \cite{goldstein2007}.
Another set of proposed measures works with fixation maps or probability distributions created by the additive superposition of Gaussians, assessing the differences between those maps employing the sum of squared pointwise subtraction \cite{wooding2002} or computing the Kullback-Leibler divergence (KLD) \cite{rajashekar2004, tatler2005}. 
Other studies computed the entropy of gaze density based on Gaussian Mixture Modeling \cite{sawahata2008, smith2013} and temporal Normalized Scanpath Saliency (NSS) \cite{Dorr.2010}, demonstrating a high correlation with KLD to quantify gaze similarity.
 A multidimensional scanpath comparison approach including gaze characteristics alongside spatial and temporal properties is the vector-based MultiMatch \cite{jarodzka2010, dewhurst2012, foulsham2012} method. The derived similarity measure combines sub-measures assessing similarity in scanpath shape, saccadic direction, length, fixation position, and duration. MultiMatch's inherent alignment of scanpaths reduces sensitivity for minor temporal shifts and manages variations in the lengths of scanpaths \cite{le2013methods}. However, a limitation of this metric is its comparison between only two scanpaths at a time, while the objective in synchrony analysis often involves comparing entire groups of subjects.

Each of the proposed methods captures slightly different characteristics of gaze behavior. For this study, we chose to compute and compare two established gaze synchrony measures that account for the properties of our relatively static video stimulus and incorporate gaze characteristics beyond spatio-temporal properties: KLD of gaze density maps and MultiMatch scanpath comparison.

\section{Methods}

\subsection{Experiment}

The ethics committee of the Faculty of Economics and Social Sciences, University of T\"ubingen (Date of approval 13 January 2022, approval \#A2.5.4-210\textunderscore ns) 
approved our study procedures, and all participants have given written consent to the data collection. 

\subsubsection{Participants}

The data employed for this study was collected from \textit{N} = 96 university students. Five participants had to be excluded from further data analysis due to technical errors during the experiment, such as malfunction of the eye tracker or crashing of experiment computers. 
Additionally, three participants were excluded from the further analysis because they were not fluent in the language used for the instructions, questionnaires, and video content. Consequently, the study was completed with a total of \textit{N} = 88 participants (Ages 19-33, \textit{M} = 23.44, \textit{SD} = 2.6), of which 19\% were male.

\subsubsection{Study Procedure and Setup}

The data was collected in the laboratory using the SMI Red remote eye tracker with a sampling rate of 250 Hz. We refrained from using chin rests, as we wanted to ensure an ecologically valid setting. Further, research has shown that even without a chin rest acceptable levels of accuracy for purposes not relying on small eye movements are achieved \cite{carter2020}. After completing a short questionnaire and a test on previous knowledge on the session topic of statistics, participants performed a nine-point pulsating calibration of the eye tracker. Participants then watched a pre-recorded Zoom lecture on introductory statistics. The video lecture's total duration was approximately 60 minutes, which required a re-calibration of the eye tracker after about 30 minutes. Participants were instructed to focus on the lecture and were not allowed to take notes or use electronic devices, including phones, during the study. After the video was completed, a comprehensive post-test of 14 questions targeting factual knowledge and deep-level understanding was conducted. 
Including the time allocated for general instructions and filling out questionnaires, the overall duration per participant averaged approximately 120 minutes. Participants received a compensation of \texteuro20.

\subsubsection{Video Stimulus}

\begin{figure}[h]
  \centering
  \includegraphics[width=0.9\linewidth]{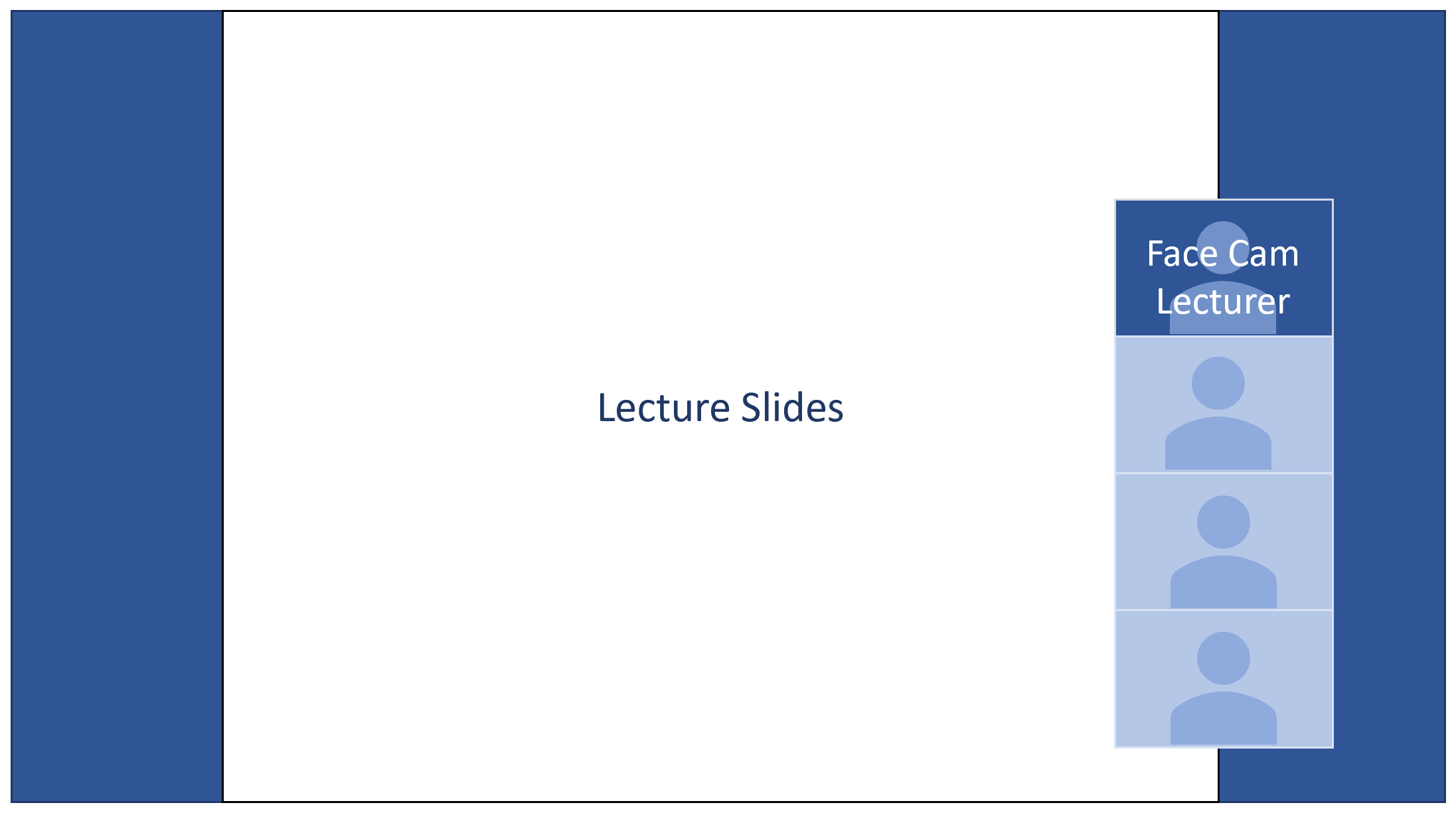}
  \caption{Zoom lecture video layout.}
  \Description{Lecture Video Layout.}
  \label{fig:vid}
\end{figure}

The video stimulus represented a typical Zoom layout, depicted in Figure \ref{fig:vid}, including lecture slides and a webcam display of the lecturer's face on the top right. Other participants in the Zoom lecture turned off their cameras so only participant tiles were visible. The slides were primarily static, yet the instructor used the cursor to point at specific locations on the slides.

\subsubsection{Attention Experience Sampling}
The most common method for directly assessing the internal state of learners is through self-reports \cite{weinstein2018}. Given the potential inaccuracies associated with retrospective reports, experience sampling, also known as probe-caught method, is typically employed. This involves intermittently stopping participants during a task and asking them to indicate where their attention was focused at that very moment. Although these self-reports are subjective, previous research has revealed their correlation with more objective but indirect measures of attention, such as physiological indicators, response times, and task performance \cite{weinstein2018}.
The study incorporated 15 quasi-randomized thought probes presented at fixed three- to five-minute intervals to assess the participants' attentional state throughout the lecture. All participants received the probes at identical moments during the lecture. A probe was administered by displaying a screen with a multiple-choice question asking what the participants thought about just now. Six answer categories adapted from \cite{Kane.2017}, ranging from ``I was on task, following the lecture'' to ``Everyday personal concerns'' (See figure \ref{fig:probe} for all categories), and an open response option was provided. Responses within the open-ended category underwent manual coding by two independent raters. The process followed an iterative method involving assigning responses to existing and establishing new off-task categories, like ``External Distraction.'' As this study focuses on the difference between attentive and inattentive learners, we dichotomized the answers accordingly. The ``I was on task, following the lecture'' answer option was coded as attentive. In contrast, answers to all other categories encompassing meta-cognitive monitoring, elaborations, distractions, and mind wandering were coded as inattentive. Although elaborations or meta-cognitive monitoring mechanisms are not inherently disruptive but are considered an essential part of the learning process, they still impede learners from following the lecture's content at that very moment.

\begin{figure}[h]
  \centering
  \includegraphics[width=1\linewidth]{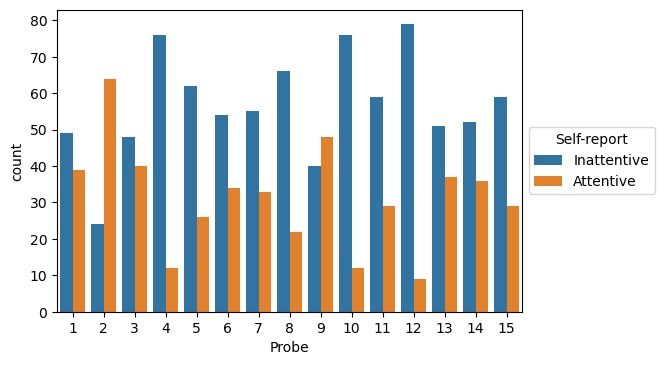}
  \caption{Absolute frequency of attention self-reports by experience sampling thought probe.}
  \Description{Barplot depicting attention self-reports by experience sampling thought probe.}
  \label{fig:attention}
\end{figure}

For $36\%$ of all thought probes during the 60-minute lecture, participants reported being on task and following the lecture. Most of the time, however, their thoughts were preoccupied with elaborations about the lecture topic, whether they understood the lecture or something else entirely, such as personal concerns or their current state, for example, tiredness or hunger ($64\%$). The difference between attentive and inattentive self-reports across all probes is significantly different from the uniform distribution ($\chi^2 = 98.337$, $p < 0.001$). Figure \ref{fig:attention} shows the absolute frequency of the participant's attention self-reports for the attentive and inattentive categories by thought probe. It becomes visible how the level of attention fluctuates over time. The peak in attention occurs at the second probe, approximately ten minutes into the lecture. Another local peak can be observed at probe nine after the eye tracker re-calibrated, which allowed for a short break.

%The high inattention levels right from the start might appear unexpected. However, many participants reported thinking about their comprehension and ideas related to the lecture content in the first four probes, coded as inattentive. Over time, thoughts that can be considered typical mind wandering, like thinking about one's current state or everyday personal concerns, occur more frequently.

\subsubsection{Learning Outcomes}

\begin{figure}[h]
  \centering
  \includegraphics[width=\linewidth]{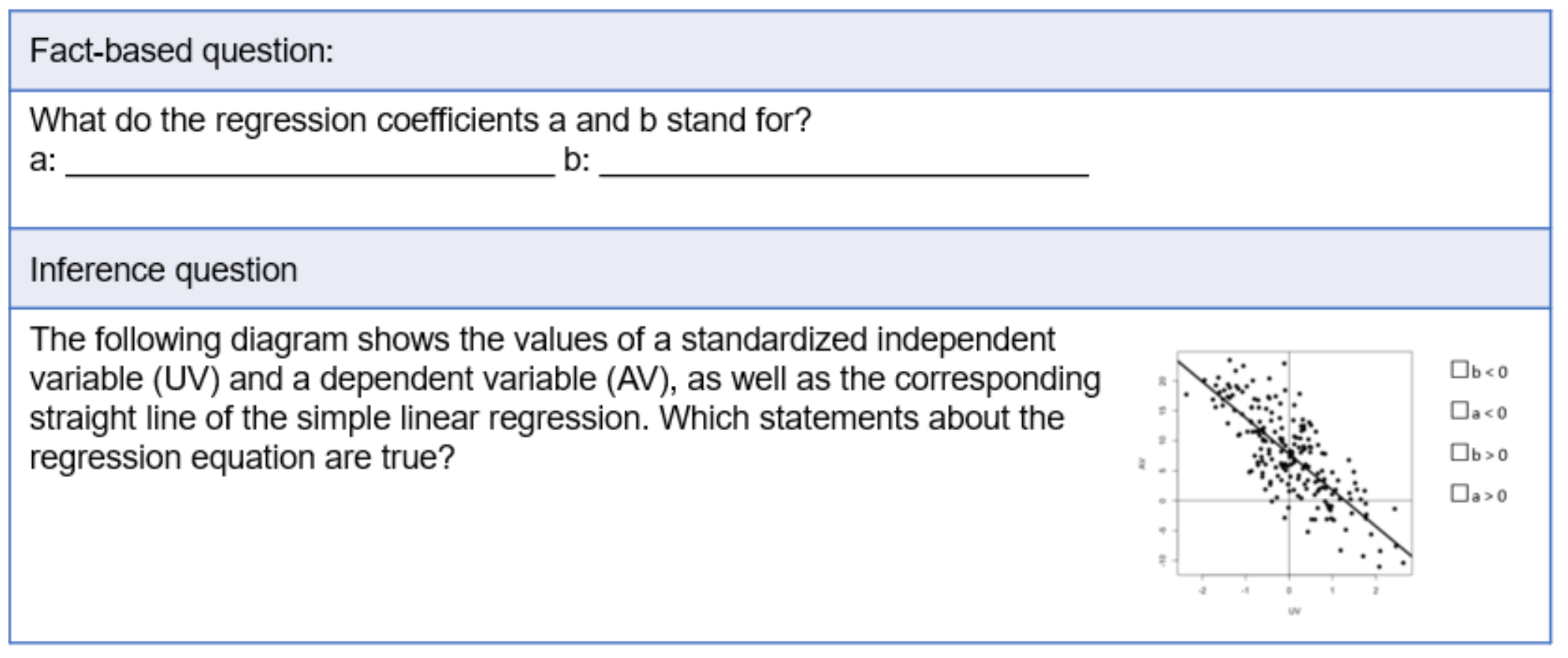}
  \caption{Example Posttest Questions.}
  \Description{Example Posttest Questions.}
  \label{fig:test}
\end{figure}

To assess learning outcomes, we administered a post-video knowledge test comprising 14 multiple-choice and open-ended questions to assess participants' comprehension levels related to the video lecture content. This assessment included seven fact-based memory questions and seven questions targeting deep-level understanding (see Figure \ref{fig:test} for an example of both types). Specifically tailored for the prerecorded lecture on linear regression analysis, the questions covered topics such as empirical covariance, method of local averaging, and least squares estimation. We ensured to include all topics covered right before the thought probes in the test. Examples of both fact-based and inference questions are illustrated in Figure \ref{fig:test}. The summative scores, ranging from 0 to 14, were derived by assigning 1 point for each correctly answered question. Participants achieved an average score of 5.63 ($\textit{sd} = 2.663$; see Figure \ref{fig:posttest} in the appendix for score distribution). To control for previous knowledge, a pre-test on the general topic of linear regression analysis was conducted before the video lecture, including eight multiple-choice and open-ended questions.

\subsection{Eye Gaze Data Pre-processing}

The average calibration error for our nine-point calibration procedure was 0.31°. We used the analysis software BeGaze by SMI to extract eye movement events, including fixations (determined by dispersion-based threshold), saccades (determined by velocity-based threshold), and blinks from the raw gaze data \cite{begaze}. We regard the time windows right before the experience sampling probes for our gaze synchrony analysis. Therefore, we cut 10-second windows before each of the 15 probes for each participant, resulting in a set of 1335 instances with corresponding self-reported attention information. Furthermore, a deficiency of the employed eye trackers that have been reported before is that tracking failures are recorded as unusually long blinks \cite{castner2020}. Consequently, blinks longer than 500 ms, exceeding an expected blink duration range between 100 to 400 ms \cite[]{schiffman2001}, were excluded. We excluded gaze sequences with less than a 75\% gaze tracking ratio to ensure high data quality. {For four participants, applying this criterion resulted in excluding all 15 sequences due to insufficient data quality with less than 75\% tracking ratio across the board. Overall, this exclusion threshold led to 785 examples from 84 participants and an average tracking ratio of 92.87\%.  

\subsection{Gaze Synchrony Assessment}

We employed two established measures to assess gaze synchrony. First, we computed the KLD between gaze density maps \cite{rajashekar2004, tatler2005}. Gaze density maps per person and regarded video sequence were created by superposing Gaussian probability density functions on fixation counts and durations of the ten-second time windows before a thought probe. When applied to gaze density maps, the KLD measures the discrepancy between two distributions of gaze points, quantifying how much one density map diverges from another \cite{le2013methods}. This involves comparing the probability distribution of gaze points across the density map generated by one set of gaze data to the reference distribution from another set, thereby indicating the degree of similarity or difference in visual attention patterns between the two. For each point on the map, the KLD quantifies the difference by calculating the logarithm of the ratio of the gaze density at that point in the first map to the gaze density at the corresponding point in the second map, then weighting this by the gaze density of the first map, and summing these values across all points. Equal gaze distributions would result in a KLD of zero, while higher values would signify a more considerable dissimilarity and, consequently, less synchronous gaze. We argue that gaze density maps, only capturing the spatial distribution of the gaze, are suitable due to the relatively static nature of the video stimulus presenting PowerPoint slides.  

Additionally, to incorporate temporal synchrony dimensions, we employed the MultiMatch scanpath comparison algorithm \cite{jarodzka2010, dewhurst2012, foulsham2012}. MultiMatch consists of five separate measures to compare scanpaths, capturing a range of characteristics: shape, direction, length, position, and duration. The shape similarity is derived by vector differences between aligned pairs of saccades and averaged over the whole scanpath. The direction subdimension is computed by the angular differences between saccades, whereas length similarity is defined as the absolute difference in amplitude of aligned saccades. These measures are insensitive to fixation locations or durations. The position similarity is computed as the Euclidean distance between aligned fixations. The duration measure is defined by the absolute difference in fixation duration of aligned fixations and is insensitive toward fixation locations and saccade information. An overall similarity score can be computed by averaging all subdimension scores. We did not simplify the scanpaths, as small changes can already be meaningful in our video setting. We used the MultiMatch\_gaze python implementation\cite{wagner2019MultiMatch} to compute similarities. To obtain an aggregated scanpath similarity measure, all five sub-measures were averaged. 

For a baseline comparison, we calculated the ISC, previously utilized in research \cite{Madsen.2021, sauter2023, Liu.2023}. This involved computing correlations for vertical and horizontal gaze positions and pupil diameter, which were then aggregated into a single ISC metric.
All synchrony measures were computed probe-wise for all 15 10-second video sequences preceding the attention self-reports separately. 
The similarity score of each participant for a given sequence was computed by comparing their gaze data with that of all other participants in the same peer group, grouped as either attentive or inattentive based on their corresponding self-reports of attention. This comparison was conducted in a pairwise fashion, and the resultant similarity scores were averaged. To compare the synchrony between the two groups across the video sequences, the synchrony values per video sequence were z-standardized.

\subsection{Analysis}

%To investigate group differences between attentive and inattentive learners we computed independent t-Tests of the computed synchrony values for both applied measures: Kullback-Leibler Distance and MultiMatch scan path comparison. For MultiMatch scan path comparison, we additionally tested for group differences in all subcategories, to investigate whether the synchrony in certain scan path characteristics have higher relevance for attentional synchrony. 
\subsubsection{Gaze Synchrony and Attention}
In the next step, we analyzed the relationship between attention self-reports and gaze synchrony. This analysis was conducted at the probe level, focusing on the relationship between each attention self-report and the gaze synchrony values computed for the 10 seconds immediately preceding the report. Since we had up to 15 probes and thus multiple measurement time points per person, we employed a multi-level analysis approach. Specifically, we utilized mixed linear regression with gaze synchrony as the dependent variable. Within this model, participant ID was treated as a random effect,  and self-reported attention was incorporated as a fixed effect. Furthermore, to account for potential variability across the video segments before each probe under analysis, we included the probe number as a categorical variable in the model. 

\subsubsection{Predicting Learning Outcomes}
To explore the potential of using gaze synchrony measures as indicators for attention, we investigated the relationship between gaze synchrony measures and learning outcomes. To this end, we aggregated gaze synchrony scores to the participant level by averaging the obtained synchrony values over all considered time windows for each participant. Similarly, to aggregate the attention self-reports, we calculated the proportion of on-task self-reports from the total number of reports for each participant, effectively determining the share of self-reported attention. With these aggregated values, we then performed linear regression analyses to compare the relation of self-reported attention, KLD, and MultiMatch similarities to post-test scores, incorporating pre-test scores into all models to adjust for prior knowledge. 

\section{Results}

%link to the results presentation: https://unitc-my.sharepoint.com/:p:/g/personal/sebea01_cloud_uni-tuebingen_de/EdJG0-4F3iFFvv8IBjpFqhUBUWkNfHXhzaoNE9BmHipBzA?e=UKlWiF

\subsection{Gaze Synchrony and Attention}

We calculated ISC as a baseline synchrony measure. When z-standar-dizing the measure at the probe level, we found average ISC values of 0.104 ($\textit{SD} = 1.019$) for the attentive group and -0.069 ($\textit{SD} = 0.983$) for the inattentive group, indicating a slightly higher gaze synchronization for attentive learners. Results of Linear mixed-effects modeling of self-reported attention on ISC can be found in the first two columns of Table \ref{tab:syncreg}. Specifically, our results indicated that participants who self-reported as attentive demonstrated significantly higher inter-subject correlation of gaze position and pupil diameter ($\textit{Estimate} = 0.21$, $\textit{p} < 0.01$) compared to those who reported being inattentive.}

\begin{table*}[h]
\caption{Linear mixed effects model of self-reported attention on Inter Subject Correlation (ISC), Kullback-Leibler divergence (KLD) and MultiMatch similarity (MM).}
\label{tab:syncreg}
\begin{adjustbox}{max width = 0.8\linewidth}
\begin{tabular}{lcccccccc}
\hline
                    & \multicolumn{2}{c}{\textbf{ISC}} & & \multicolumn{2}{c}{\textbf{KLD}}       &         & \multicolumn{2}{c}{\textbf{MM}} \\
Predictors          & \multicolumn{1}{c}{\textit{Estimates}} & \multicolumn{1}{c}{\textit{CI}}  && \multicolumn{1}{c}{\textit{Estimates}}               & \multicolumn{1}{c}{\textit{CI}}           & & \multicolumn{1}{c}{\textit{Estimates}}       & \multicolumn{1}{c}{\textit{CI}}            \\ \hline
(Intercept)         & -0.09              & -0.34 – 0.15 && 0.17                    & -0.07 – 0.41  && -0.21           & -0.45 – 0.03  \\
Attentive           & 0.21 **            & 0.06 – 0.37  && -0.36 ***               & -0.51 – -0.22 && 0.38 ***        & 0.23 – 0.53   \\
Stimulus 2          & -0.06              & -0.39 – 0.27 && 0.13                    & -0.17 – 0.43  && -0.08           & -0.39 – 0.22  \\
Stimulus 3          & -0.02              & -0.37 – 0.33 && 0.12                    & -0.20 – 0.43  && -0.04           & -0.36 – 0.28  \\
Stimulus 4          & 0.06               & -0.28 – 0.39 && -0.08                   & -0.38 – 0.21  && 0.10            & -0.20 – 0.41  \\
Stimulus 5          & 0.04               & -0.31 – 0.39 && -0.03                   & -0.35 – 0.29  && 0.09            & -0.24 – 0.41  \\
Stimulus 6          & -0.01              & -0.38 – 0.35 && 0.08                    & -0.25 – 0.40  && -0.02           & -0.36 – 0.31  \\
Stimulus 7          & 0.01               & -0.34 – 0.37 && 0.09                    & -0.23 – 0.41  && 0.01            & -0.32 – 0.34  \\
Stimulus 8          & 0.04               & -0.32 – 0.40 && 0.02                    & -0.30 – 0.35  && 0.04            & -0.30 – 0.37  \\
Stimulus 9          & -0.04              & -0.40 – 0.32 && 0.13                    & -0.19 – 0.45  && -0.09           & -0.42 – 0.24  \\
Stimulus 10         & 0.07               & -0.31 – 0.45 && -0.08                   & -0.42 – 0.26  && 0.09            & -0.26 – 0.44  \\
Stimulus 11         & 0.02               & -0.35 – 0.38 && 0.06                    & -0.27 – 0.39  && -0.01           & -0.35 – 0.33  \\
Stimulus 12         & 0.07               & -0.29 – 0.44 && -0.18                   & -0.51 – 0.15  && 0.19            & -0.15 – 0.54  \\
Stimulus 13         & -0.02              & -0.40 – 0.36 && 0.05                    & -0.29 – 0.39  && -0.04           & -0.39 – 0.31  \\
Stimulus 14         & -0.01              & -0.40 – 0.37 && 0.04                    & -0.30 – 0.39  && -0.02           & -0.38 – 0.34  \\
Stimulus 15         & 0.02               & -0.38 – 0.42 && -0.02                   & -0.38 – 0.34  && 0.01            & -0.36 – 0.38  \\
Random Effects      &                    &              &&                         &               &&                 &               \\
$\sigma^2$          & 1                  &              && 0.79                    &               && 0.84            &               \\
$\tau_{00id}$              & 0.02               &              && 0.21                    &               && 0.17            &               \\
ICC                 & 0.02               &              && 0.21                    &               && 0.17            &               \\
$N_{id}$                 & 84                 &              && 84                      &               && 84              &               \\ \hline
Observations        & 785                &              && 785                     &               && 785             &               \\
Marg. R2 / Cond. R2 & \multicolumn{2}{l}{0.009   / 0.025} & &  \multicolumn{2}{l}{0.028 / 0.234} && \multicolumn{2}{l}{0.029 / 0.196 } \\ \hline
\multicolumn{9}{r}{* p\textless{}0.05   ** p\textless{}0.01   *** p\textless{}0.001}                                               
\end{tabular}
\end{adjustbox}
\end{table*}

\subsubsection{Kullback-Leibler Divergence}

%video source: https://www.youtube.com/watch?v=7H_DUs3ENaQ
%ttest: statistic=5.036885834788937, pvalue=5.87762416626359e-07

\begin{figure}[]
  \centering
  \includegraphics[width=1\linewidth]{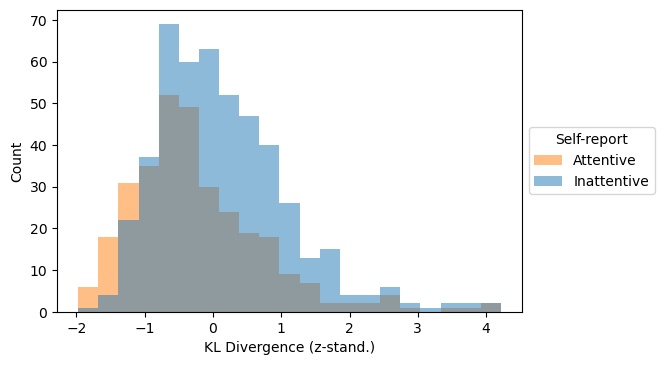}
  \caption{Kullback-Leibler divergence of gaze density maps by self-reported attention, z-standardized by video sequence.}
  \Description{Histogroam depicting Kullback-Leibler Divergence of Gaze Density Maps by Self-Reported Attention, Z-Standardized by Video Sequence.}
  \label{fig:kldiv}
\end{figure}

The analysis of gaze density maps, utilizing KLD to quantify synchrony in gaze patterns, was conducted probe-wise. Each video sequence, spanning 10 seconds and leading up to a self-report probe, was individually z-standardized to ensure comparability across different segments. The standardization process modifies the KLD scale, allowing us to interpret smaller or negative values as indicative of reduced differences between gaze maps, thereby signifying greater gaze synchrony. The distribution of KLD, divided into attentive ($\textit{M} = -0.219$, $\textit{SD} = 1.009$) and inattentive ($\textit{M} = 0.142$, $\textit{SD} = 0.968$) instances based on participant's self-reports are depicted in Figure \ref{fig:kldiv}. The visible delineation suggests that participants who self-reported as inattentive exhibited a marginally higher divergence in gaze patterns within their peer group, indicating a reduced level of gaze synchrony compared to their attentive counterparts. This higher divergence signifies smaller gaze synchrony within the distracted group. Although the trend is visible, the two distributions largely overlap, illustrating that the differences are not particularly large.

%The performed multi-level analysis revealed that participants who reported being attentive showed significantly lower gaze divergence ($\textit{Estimate} = -0.36$, $\textit{p} < 0.001$) than participants reporting being distracted, synonymous with higher gaze synchrony (See appendix Table \ref{tab:KLDreg} for the full regression table). 

Further exploration through multi-level analysis, employing linear mixed effects models, accentuated these findings. Specifically, our results indicated that participants who self-reported as attentive demonstrated significantly lower gaze divergence ($\textit{Estimate} = -0.36$, $\textit{p} < 0.001$) compared to those who reported being inattentive. This statistical significance underscores a greater degree of gaze synchrony among participants who were on task and focused on the lecture, as per their self-reports, albeit by a small effect size. Such a relationship between self-reported attention states and gaze synchrony metrics, detailed in Table \ref{tab:syncreg}, provides empirical evidence supporting the assumption that attention levels, as self-reported by participants, are intricately associated with measurable gaze behaviors during video lecture viewing.

\begin{figure*}[h]
  \centering
  \includegraphics[width=\linewidth]{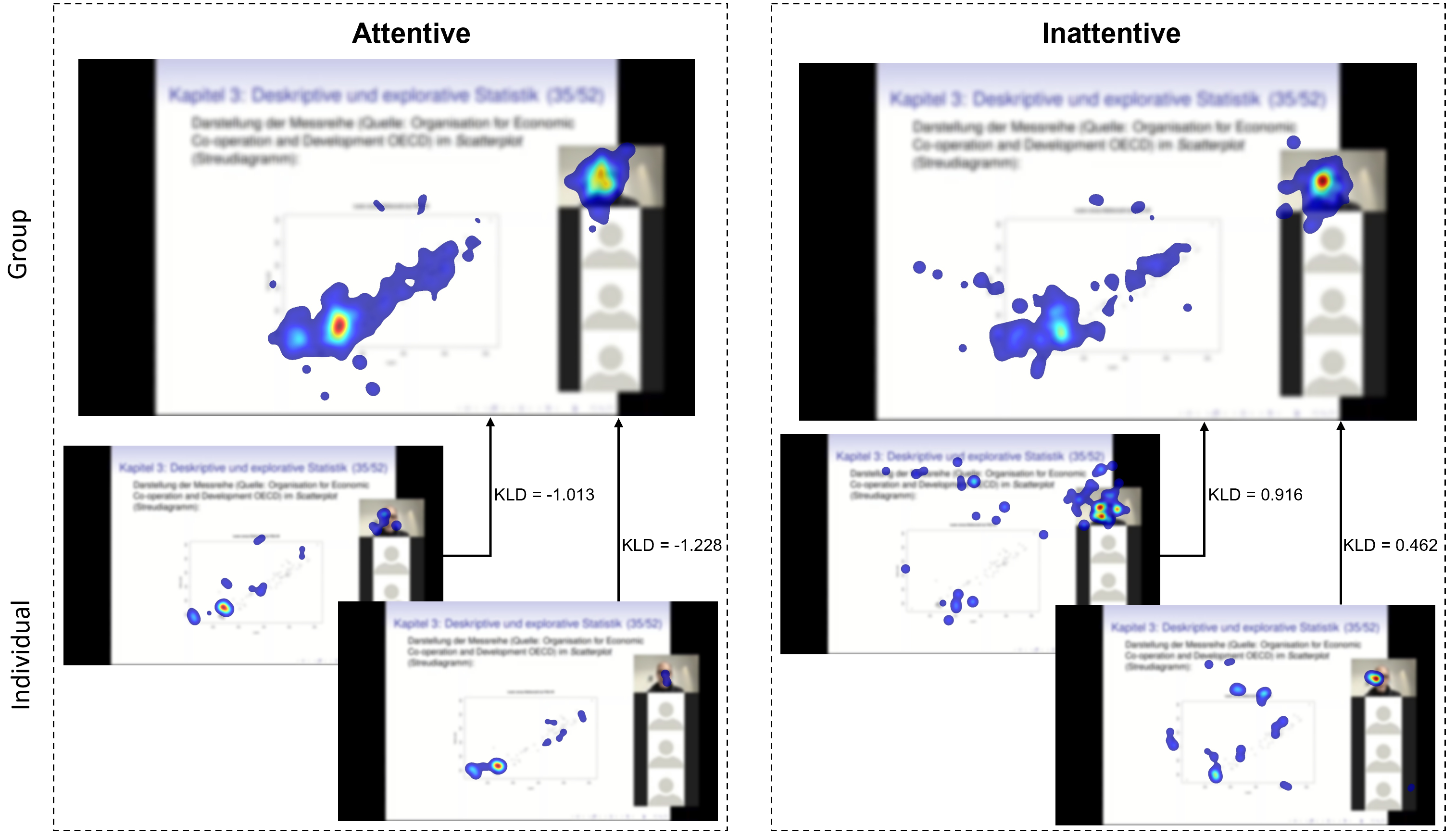}
  \caption{Example gaze density heatmap visualizations by attentive and inattentive self-reports in a video sequence. The top row shows the gaze density at the group level, while images on the bottom show examples of individual participants' gazes and corresponding average Kullback-Leibler Divergences (KLD).}
  \Description{}
  \label{fig:hm}
\end{figure*}

Figure \ref{fig:hm} shows visualizations of gaze density maps for attentive and inattentive learners, according to their self-reports, for one video sequence of 10 seconds before a preceding probe. The group-level depiction of gaze density illustrates that the attentive group focused clearly on a specific point in the graphic on the slide shown. In contrast, the gaze of the distracted group tended to be scattered across the slide, and a high gaze density can only be seen directly by the lecturer. When looking at examples of single participants, attentive heatmaps appeared more similar to each other and to the group depiction, which is supported by the lower KLD values of $-1.013$ and $-1.228$. Contrarily, the inattentive participants' gaze patterns appeared less focused and more random, also reflected in higher KLD Values of $0.916$ and $0.462$. Additionally, a correlation analysis between the baseline ISC similarity scores and KLD distance values revealed a weak negative significant correlation ($\textit{r} = -0,12$).

%The performed independent t-Test revealed, that in the 313 instances in which participants reported being on task, following the lecture, ($\textit{M} = -0.219$, $\textit{SD} = 1.009$) compared to the 472 instances in which participants reported thinking about something else ($\textit{M} = 0.142$, $\textit{SD} = 0.968$) participants demonstrated significantly lower gaze divergence from the group of attentive participants, synonymous with higher gaze synchronicity, $\textit{t} = 5.037$, $\textit{p} = 5.87e-07$.

% include gaze density heatmaps of both groups and single participants

\subsubsection{Scanpath Comparisons with MultiMatch}

\begin{figure}[h]
  \centering
  \includegraphics[width=1\linewidth]{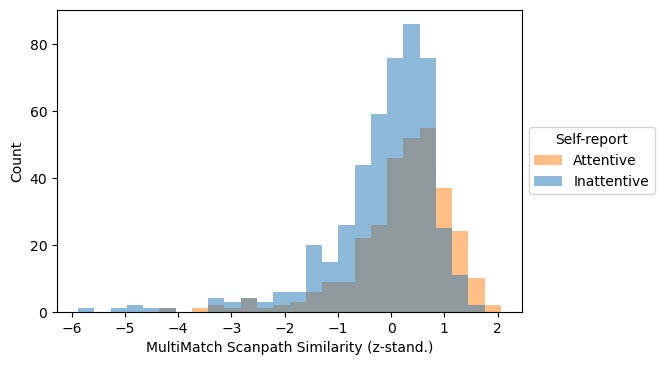}
  \caption{MultiMatch scanpath similarity scores by self-reported attention, z-standardized by video sequence.}
  \Description{Histogroam depicting MultiMatch Scanpath Similarity Scores by Self-Reported Attention, Z-Standardized by Video Sequence.}
  \label{fig:mmsim}
\end{figure}

As a second measure of synchrony, we employed the MultiMatch method to assess scanpath similarities computed at the probe level for individual instances. 
The calculated similarities are presented in Figure \ref{fig:mmsim}, where we observed that attentive participants ($\textit{M} = 0.207$, $\textit{SD} = 0.946$) exhibited marginally more similar scanpaths compared to their inattentive counterparts ($\textit{M} = -0.137$, $\textit{SD} = 1.013$). This suggests a higher level of gaze synchronization among participants who reported being on task during the respective video lecture sequences.

%The distribution of gaze similarities obtained by the multidimensional MultiMatch method to compute scanpath similarities between attentive ($\textit{M} = 0.207$, $\textit{SD} = 0.946$) and inattentive ($\textit{M} = -0.137$, $\textit{SD} = 1.013$) participants can be seen in Figure \ref{fig:mmsim}. Attentive participants show slightly more similar scanpaths than distracted participants. 

%When investigating group differences with mixed linear models, we found that attentive learners show significantly higher MM scanpath similarities ($\textit{Estimate} = 0.38, \textit{p} < 0.001$; See Table  in the appendix for a full regression table) and therefore higher gaze synchronization. We found significant differences regarding attention self-report for all subdimensions of MultiMatch, except for the similarity of saccade lengths ($\textit{Estimate} = 0.02, \textit{p} = 0.825$; See Table~\ref{tab:MMreg}). The strongest effect of attention was found on gaze position similarity ($\textit{Estimate} = 0.42, \textit{p} < 0.001$). Additionally, we compared the MultiMatch similarities to the previously computed KLD on an instance level. MultiMatch and KLD show a small, significant correlation of $\textit{r} = -0.45$.

To delve deeper into the differences between groups, we applied mixed linear models to analyze the data at the probe level. Table \ref{tab:syncreg} displays the models for MM and its subdimensions. This analysis revealed that attentive learners demonstrated higher MM scanpath similarities ($\textit{Estimate} = 0.38, \textit{p} < 0.001$), indicating a greater degree of gaze synchronization though the effect size suggests these differences, while statistically significant, are modest in magnitude. This finding was consistent across all subdimensions of the MM analysis, as displayed in Table \ref{tab:MMreg}, with the exception of saccade length similarity ($\textit{Estimate} = 0.02, \textit{p} = 0.825$), which did not show a significant difference between attentive and inattentive groups. Notably, the most pronounced effect of attention was observed in the similarity of gaze positions ($\textit{Estimate} = 0.42, \textit{p} < 0.001$), underscoring the impact of the attentional state on visual engagement with the video content.
When comparing KLD scores to our ISC baseline, we found a weak, significantly positive correlation ($\textit{r} = 0,17$). Further, we conducted a comparative analysis of MM similarities and KLD scores. This comparison revealed a modest yet significant correlation ($\textit{r} = -0.45$) between MultiMatch and KLD, suggesting that both metrics, although distinct in their computational approaches, provide complementary insights into the nature of gaze behavior and its association with attention.

%When statistically comparing the groups with an independent T-Test, it reveals that in the 313 instances in which participants paid attention to the lecture ($\textit{M} = 0.207$, $\textit{SD} = 0.946$) compared to the 472 sequences in which participants reported to be distracted ($\textit{M} = -0.137$, $\textit{SD} = 1.013$) scanpath similarities were significantly higher, $\textit{t} = 4.786$, $\textit{p} = 2.038e-06$. Accordingly, also when comparing spatial and time sensitive scanpaths attentive learners show a higher gaze synchrony.

% only subdimension no significant difference: length

\begin{figure*}[h]
  \centering
  \includegraphics[width=0.8\linewidth]{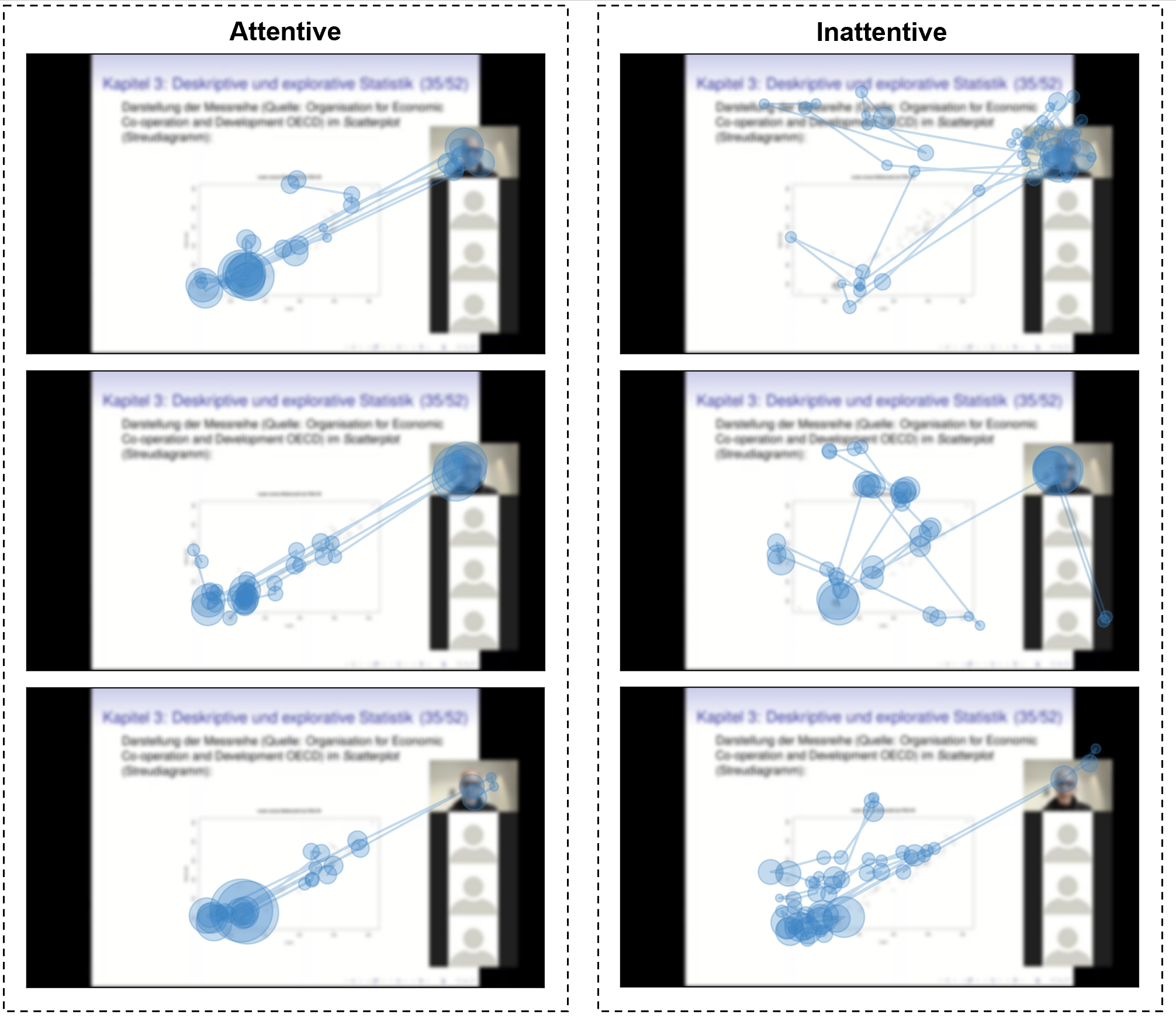}
  \caption{Example scannpath visualizations of one 10-second video sequence by self-reported attention.}
  \label{fig:sp}
\end{figure*}

A visualization of example scanpaths can be seen in Figure \ref{fig:sp}. The three participants depicted, attentive according to their self-reports, show very similar fixation patterns that move back and forth between the presenter and the very few specific relevant points on the slides. The scanpaths of distracted learners exhibit significantly greater diversity. In these cases, fixations are dispersed across a broader area of the slide and do not seem to follow a distinct pattern. While their gaze also briefly touches the relevant areas of the graphic, they occasionally fixate on empty areas on the slides.

%T-Test: statistic=4.785531542481849, pvalue=2.0384945830232095e-06

\subsection{Predicting Learning Outcomes}

We explored the relationship between gaze synchrony as an indicator of attention and learning outcomes by aggregating self-reported attention and gaze synchrony metrics to the participant level and then conducting linear regression analysis on post-test scores. This approach facilitates a detailed exploration and comparison of the relationships between self-reported attention and gaze synchrony with learning outcomes, examining how each correlates with educational success over the entire session. 
The results of linear regression analysis are detailed in Table \ref{tab:regpt}. Our findings indicate a significant positive relationship between the overall proportion of time participants reported being on task and their post-test scores, even after adjusting for their prior knowledge of the session topic. This suggests that self-reported attention strongly predicts learning success, underscoring the importance of maintaining focus during educational sessions.
In contrast, we could not find a significant relationship between the computed gaze (a-)synchrony measures KLD and MultiMatch similarity and learning outcomes. However, in these models, previous knowledge becomes significant. This lack of significant correlation suggests that while these measures provide valuable insights into participants' engagement and attention alignment, they may not directly predict learning effectiveness as measured by post-test scores. Interestingly, in the models incorporating gaze synchrony metrics, prior knowledge emerged as a significant predictor. Models that do not account for previous knowledge are detailed in the appendix, specifically in Table \ref{tab:singlept_reg}.

\begin{table*}[h]
\caption{Linear regression of share of attentive self-reports, average Kullback-Leibler divergence (KLD), and average MultiMatch scanpath similarity (MM) on posttest scores, controlling for pretest scores.}
\label{tab:regpt}
\begin{tabular}{lllllll}
\hline
                                                                      & \multicolumn{2}{l}{}              & \multicolumn{2}{c}{Post-Test Score} & \multicolumn{2}{l}{}              \\
Predictors                                                            & Estimates      & CI               & Estimates       & CI                & Estimates      & CI               \\ \hline
Intercept                                                             & $2.89^{\ast\ast\ast}$      & 1.63 – 4.15      & $5.16^{\ast\ast\ast}$       & 4.44 – 5.88       & $5.17^{\ast\ast\ast}$      & 4.46 – 5.88      \\
Pre-Test Score                      & 0.55           & -0.06 – 1.15     & $0.72^{\ast}$         & 0.05 – 1.39       & $0.70^{\ast}$        & 0.03 – 1.37      \\
Attentive Share & $0.05^{\ast\ast\ast}$     & 0.03 – 0.08      &                 &                   &                &                  \\
Average KLD                                                           &                &                  & 0.14            & -0.74 – 1.02      &                &                  \\
Average MM                                                            &                &                  &                 &                   & -0.17          & -1.06 – 0.72     \\ \hline
Observations                                                          & \multicolumn{2}{l}{84}            & \multicolumn{2}{l}{84}              & \multicolumn{2}{l}{84}            \\
R2 / R2 adjusted                                                      & \multicolumn{2}{l}{0.224 / 0.205} & \multicolumn{2}{l}{0.054 / 0.030}   & \multicolumn{2}{l}{0.054 / 0.031} \\ \hline
\multicolumn{7}{r}{ $^{\ast}$ p\textless{}0.05    $^{\ast\ast}$ p\textless{}0.01    $^{\ast\ast\ast}$ p\textless{}0.001}                                                                                               
\end{tabular}
\end{table*}

\section{Discussion}

Our study fills a crucial gap by exploring the relationship between gaze synchrony and self-reported attention during lecture video watching. We identified significant differences in gaze synchrony by self-reported attention, indicating higher synchronization when students report attentiveness. However, these differences were observed to be relatively small in magnitude. As a first study, this work establishes a connection between gaze synchrony and experience-sampled attention reports, reinforcing the hypothesis that gaze synchrony, beyond experimental conditions, is related to naturally occurring (in)attentiveness during video lecture viewing.

When comparing and contrasting the two measures employed to assess gaze synchrony, namely the Kullback-Leibler divergence and the MultiMatch Scanpath comparison, the small significant correlation reveals a common trend but shows that the two measures still depict distinct characteristics of the eye movements. While KLD focuses mainly on the spatial distribution of fixations, MultiMatch incorporates the temporal dimension by considering the sequence and a range of other multidimensional gaze properties, such as overall scanpath shape. Interestingly, the sub-dimension of MultiMatch that is most strongly linked to self-reported attention is the one that assesses how similar fixation positions are. This underscores the significance of where the eyes focus and the visual engagement with specific content.
On the contrary, the only subdimension that did not show a significant relation was saccade length similarity. This discrepancy may indicate that the amplitude of aligned eye movements does not exhibit increased synchronization in the same way as other aspects of gaze behavior when learners are attentively engaged in a task. Location may be more directly influenced by where attention is focused, reflecting the cognitive engagement with specific content areas.
Consequently, both measures appear to be suitable for assessing gaze synchrony in the setting of video lecture learning while providing complementary insights. When synchrony is assessed with these measures, it demonstrates a stronger connection to self-reported attention than the previously used ISC measure. This likely stems from their lower time sensitivity, which is more suitable for relatively static stimuli like lecture slides. The exemplary visualization of gaze density heatmaps and scanpaths shows that in many cases, the discrepancy in gaze movement patterns associated with different attentional states can be readily discerned through visual inspection, as described in previous studies \cite{Sauter.2022b}.

Our study did not replicate the previously suggested finding \cite{Madsen.2021, Liu.2023} that the average level of gaze synchrony significantly predicts post-test scores. Conversely, as anticipated, self-reported attention demonstrated a significant association with learning outcomes. A similar finding was reported by \cite{Sauter.2022}, who attributed the lack of association between gaze synchrony and test scores primarily to the poorer webcam-based eye-tracking sampling rates. However, these diverging findings may also partially be attributed to distinctions between the employed video stimuli. \cite{Madsen.2021} used short, informal instructional videos, i.e., animations, while \cite{Sauter.2022} and this study displayed more traditional lecture videos featuring slides and a lecturer. These differences in video content might influence the observed gaze synchrony patterns, highlighting the potential impact of video format on eye movement behaviors. The presence of a presenter in the video changes the gaze distribution and potentially how eye movements synchronize during attention. This suggestion is also supported by findings of \cite{Sauter.2022}, that the amount of gaze on the presenter interacted with ISC in the regression on post-test scores. Other studies revealed longer fixations on the instructor during mind-wandering episodes \cite{zhang2020}.

A limitation of this study is the observation of relatively short time windows for calculating gaze synchrony instead of the continuous observation over the entire lecture video. These brief intervals were specifically chosen to accommodate the inherently fluctuating nature of attention, recognizing that it can vary significantly within short periods. However, this methodological choice means that only a small proportion of the total gaze data is utilized for synchrony computation and, consequently, to predict learning outcomes. Observing eye movements over longer periods could increase the synchrony measures' robustness and potentially reveal a clearer relation to learning outcomes. 
Further, our study sample consists of university students within a narrow age range, potentially affecting the generalizability to other age groups such as school children. Moreover, the data is unbalanced in terms of gender, with a smaller proportion of male participants. Additionally, the chosen video lecture, which is not part of participants' regular study programs, may have contributed to lower intrinsic motivation, influencing their attention. While our findings provide important insights into gaze synchrony and attention, they may not directly apply to live online lectures, where participants might be visible through a webcam. Prior research has identified a negative correlation between the time spent actively looking at one's own and other students' webcam images and learning outcomes \cite{wisiecka2022}. This suggests that the mere visibility of these images could act as a distractor in live online educational contexts. Another limitation of the current study is that in our effort to increase ecological validity, we refrained from using chin rests. This decision, aimed at creating a more naturalistic setting for participants, might have affected the accuracy of our eye-tracking measurements.

%Add limitations (Window size, synchrony scores over whole video)

%future research synchro

The observations in this study underline the complexity of directly using synchrony as an attention indicator, suggesting that the findings are not as robust as previously thought. Future research should investigate how gaze synchrony is influenced by the educational video type, especially by the presence of a presenter. The choice of a synchrony measure in future research should be informed by the dynamic nature of the video or learning task at hand. Considering that KLD and MM appear to perform comparably for static, slide-based stimuli, their selection might hinge on the computational demands for real-time applications. KLD might be more computationally efficient, especially for group-based comparisons. Nonetheless, for dynamic stimuli or in scenarios involving longer time windows, MM might be preferable due to its capacity to incorporate the temporal dimension. Additionally, the moderate correlation observed between KLD and MM suggests they could be complementary. To leverage the unique insights provided by each, employing a hybrid approach, such as Normalized Scanpath Saliency, may offer significant benefits \cite{le2013methods, Dorr.2010}. Future research should systematically test and compare various synchrony measures and how they relate to video types to find a more precise measure of synchrony that can eventually serve as a robust attention indicator during online learning. 

This is particularly relevant for further research aimed at a better understanding of the learning process in online environments and, for example, improving the quality of learning materials. Moreover, this advancement carries significant implications for practical applications. Prior research proposed the potential of employing webcam-based eye tracking to enable real-time adaptability in online education based on attention \cite{Madsen.2021}. The finding that gaze synchrony correlates with momentary, naturally occurring attention presents the prospect that an overall degree of gaze synchrony could be a meaningful metric for lecturers and online educators, providing insights into the attentiveness of learners in live online settings. This could potentially guide instructional strategies to enhance student engagement and learning outcomes. However, ethical considerations take center stage when contemplating the application of eye tracking in real-world settings. Researchers must prioritize students' privacy \cite{suemer_2020_automated_anon} by implementing robust measures to secure and anonymize eye movement data~\cite{diff_privacy_eye_tracking_2021, brendan_streaming_et_data_tvcg_2021}, as using eye movement data, it is possible to infer various sensitive user attributes~\cite{Liebling_preibusch_2014, Kroeger_etal_2020}. Therefore, it is also essential to carefully consider issues of consent, data security, and the responsible use of technology. Furthermore, researchers should be mindful of the potential impact on learners, and steps should be taken to avoid potential biases that can disadvantage learner groups. 

\section{Conclusion}

In conclusion, this study investigated the relationship between gaze synchrony and self-reported attention in a realistic video lecture setting. While we found attentive participants exhibited higher synchronization of eye movements, our results did not show a significant association between gaze synchrony and learning outcomes. The findings underscore the complexity of using gaze synchrony as a reliable indicator of attention. Further research is required to explore the interplay between attention, gaze synchrony, and the educational video type to better understand their relationship.

\bibliographystyle{ACM-Reference-Format}
\bibliography{A_Lit}

\begin{acks}

This research was supported by the LEAD Graduate School \& Research Network, which is funded by the Ministry of Science, Research and the Arts of the state of Baden- W{\"u}rttemberg within the framework of the sustainability funding for the projects of the Excellence Initiative II. Babette B{\"u}hler and Hannah Deininger are doctoral candidates supported by the LEAD Graduate School and Research Network. Efe Bozkir and Enkelejda Kasneci acknowledge the funding by the Deutsche Forschungsgemeinschaft (DFG, German Research Foundation) under Germany's Excellence Strategy - EXC number 2064/1 - Project number 390727645.

\end{acks}

\appendix
\section{Appendices}

\begin{figure}[h]
  \centering
  \includegraphics[width=\linewidth]{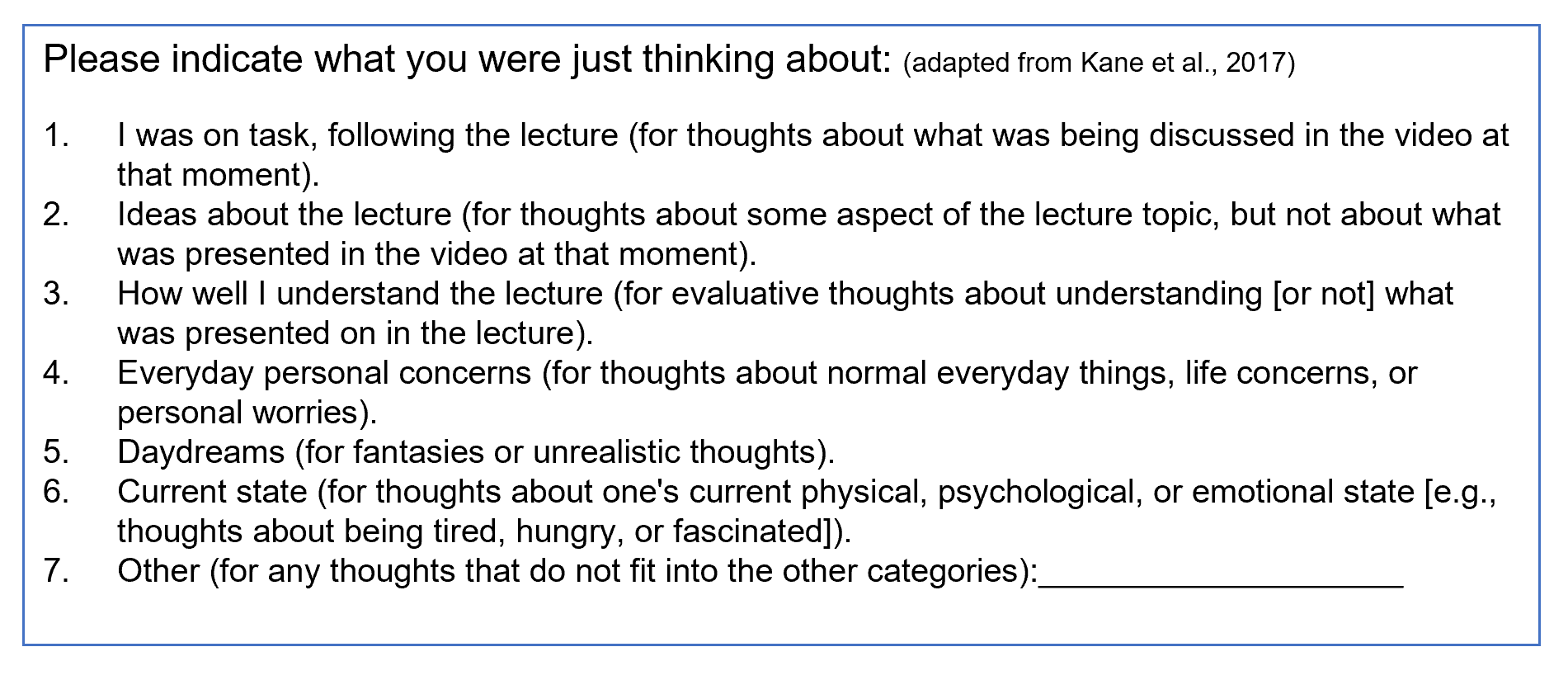}
  \caption{Attention thought probes, adapted from \cite{Kane.2017}.}
  \Description{}
  \label{fig:probe}
\end{figure}

\begin{figure}[h]
  \centering
  \includegraphics[width=0.8\linewidth]{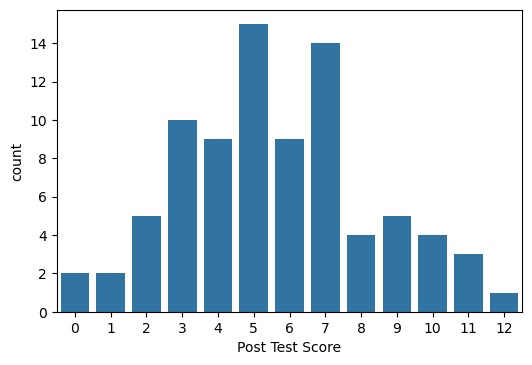}
  \caption{Post test scores.}
  \Description{}
  \label{fig:posttest}
\end{figure}

\begin{table*}[]
\caption{Linear mixed effects models of self-reported attention on MultiMatch scanpath similarity (MM) sub-dimensions.}
\label{tab:MMreg}
\begin{adjustbox}{max width = \linewidth}
\begin{tabular}{lcccccccccc}
\cline{1-11}
                                                                                     & \multicolumn{2}{c}{\textbf{MM Shape}}      & \multicolumn{2}{c}{\textbf{MM Length}}     & \multicolumn{2}{c}{\textbf{MM Direction}}  & \multicolumn{2}{c}{\textbf{MM Position}}   & \multicolumn{2}{c}{\textbf{MM Duration}}    \\
Predictors                                                                           & \textit{Estimates}      \textit{CI}               & \textit{Estimates}      & \textit{CI}               & \textit{Estimates}      & \textit{CI}               & \textit{Estimates}      & \textit{CI}               & \textit{Estimates}      & \textit{CI}                 \\ \hline
(Intercept)                                                                          & -0.28 *        & -0.54 – -0.02    & -0.03          & -0.27 – 0.21     & -0.30 *        & -0.57 – -0.04    & -0.20          & -0.44 – 0.04     & -0.16          & -0.40 – 0.08       \\
Attentive                                                                            & 0.30 ***       & 0.17 – 0.42      & 0.02           & -0.14 – 0.17     & 0.29 ***       & 0.17 – 0.42      & 0.42 ***       & 0.27 – 0.58      & 0.30 ***       & 0.15 – 0.45        \\
Stimulus 2                                                                           & -0.05          & -0.30 – 0.19     & 0.01           & -0.31 – 0.33     & -0.05          & -0.30 – 0.20     & -0.12          & -0.44 – 0.21     & -0.06          & -0.37 – 0.25       \\
Stimulus 3                                                                           & -0.02          & -0.27 – 0.24     & 0.01           & -0.32 – 0.35     & -0.02          & -0.29 – 0.24     & -0.05          & -0.39 – 0.28     & -0.03          & -0.35 – 0.30       \\
Stimulus 4                                                                           & 0.05           & -0.20 – 0.29     & 0.02           & -0.30 – 0.34     & 0.06           & -0.20 – 0.31     & 0.12           & -0.21 – 0.44     & 0.09           & -0.23 – 0.40       \\
Stimulus 5                                                                           & 0.06           & -0.20 – 0.32     & 0.04           & -0.30 – 0.38     & 0.06           & -0.21 – 0.33     & 0.08           & -0.27 – 0.42     & 0.07           & -0.27 – 0.40       \\
Stimulus 6                                                                           & -0.04          & -0.30 – 0.23     & 0.01           & -0.34 – 0.36     & -0.03          & -0.30 – 0.25     & -0.03          & -0.38 – 0.32     & -0.01          & -0.35 – 0.33       \\
Stimulus 7                                                                           & -0.04          & -0.30 – 0.23     & 0.01           & -0.33 – 0.36     & -0.02          & -0.29 – 0.25     & -0.00          & -0.35 – 0.35     & 0.03           & -0.30 – 0.37       \\
Stimulus 8                                                                           & -0.01          & -0.28 – 0.25     & 0.00           & -0.34 – 0.35     & -0.01          & -0.29 – 0.26     & 0.06           & -0.29 – 0.41     & 0.04           & -0.30 – 0.38       \\
Stimulus 9                                                                           & -0.02          & -0.29 – 0.24     & -0.01          & -0.36 – 0.33     & -0.03          & -0.30 – 0.25     & -0.07          & -0.42 – 0.27     & -0.07          & -0.41 – 0.27       \\
Stimulus 10                                                                          & 0.17           & -0.11 – 0.45     & -0.02          & -0.38 – 0.34     & 0.16           & -0.13 – 0.45     & 0.14           & -0.23 – 0.50     & 0.07           & -0.29 – 0.42       \\
Stimulus 11                                                                          & 0.05           & -0.22 – 0.33     & -0.02          & -0.37 – 0.33     & 0.05           & -0.23 – 0.33     & 0.02           & -0.34 – 0.37     & 0.00           & -0.34 – 0.35       \\
Stimulus 12                                                                          & 0.26           & -0.02 – 0.53     & 0.03           & -0.33 – 0.38     & 0.25           & -0.03 – 0.53     & 0.18           & -0.18 – 0.54     & 0.13           & -0.22 – 0.48       \\
Stimulus 13                                                                          & 0.13           & -0.15 – 0.41     & -0.04          & -0.40 – 0.33     & 0.11           & -0.18 – 0.40     & -0.05          & -0.41 – 0.32     & -0.04          & -0.40 – 0.32       \\
Stimulus 14                                                                          & 0.02           & -0.26 – 0.31     & 0.02           & -0.35 – 0.39     & 0.04           & -0.25 – 0.33     & -0.02          & -0.39 – 0.35     & -0.02          & -0.38 – 0.34       \\
Stimulus 15                                                                          & 0.09           & -0.21 – 0.38     & 0.00           & -0.38 – 0.39     & 0.08           & -0.22 – 0.39     & 0.03           & -0.36 – 0.41     & 0.00           & -0.38 – 0.38       \\
Random Effects                                                                       &                &                  &                &                  &                &                  &                &                  &                &                    \\
$\sigma^2$                                                                                     & \multicolumn{2}{l}{0.52}          & \multicolumn{2}{l}{0.90}          & \multicolumn{2}{l}{0.56}          & \multicolumn{2}{l}{0.93}          & \multicolumn{2}{l}{0.87}            \\
$\tau_{00id}$                                                                                  & \multicolumn{2}{l}{0.77}       & \multicolumn{2}{l}{0.12}       & \multicolumn{2}{l}{0.78}       & \multicolumn{2}{l}{0.05}       & \multicolumn{2}{l}{0.14}         \\
ICC                                                                                  & \multicolumn{2}{l}{0.60}          & \multicolumn{2}{l}{0.12}          & \multicolumn{2}{l}{0.58}          & \multicolumn{2}{l}{0.05}          & \multicolumn{2}{l}{0.14}            \\
$N_{id}$                                                                                    & \multicolumn{2}{l}{84}         & \multicolumn{2}{l}{84}         & \multicolumn{2}{l}{84}         & \multicolumn{2}{l}{84}         & \multicolumn{2}{l}{84}           \\ \hline
Observations                                                                         & \multicolumn{2}{l}{785}           & \multicolumn{2}{l}{785}           & \multicolumn{2}{l}{785}           & \multicolumn{2}{l}{785}           & \multicolumn{2}{l}{785}             \\
Marg. R2 / Cond. R2                                                         & \multicolumn{2}{l}{0.017 / 0.603} & \multicolumn{2}{l}{0.000 / 0.122} & \multicolumn{2}{l}{0.015 / 0.589} & \multicolumn{2}{l}{0.036 / 0.087} & \multicolumn{2}{l}{0.018 / 0.153}   \\ \hline \multicolumn{11}{r}{* p\textless{}0.05   ** p\textless{}0.01   *** p\textless{}0.001} 
\end{tabular}
\end{adjustbox}
\end{table*}

\begin{table*}[h]
\caption{Linear regression of pre-test Score, attentive self-report share, average Kullback-Leibler divergence (KLD) and average MultiMatch scanpath similarity (MM) on posttest scores.}
\label{tab:singlept_reg}
\begin{adjustbox}{max width = \linewidth}
\begin{tabular}{lllllllll}
\hline
                            & \multicolumn{8}{c}{\textbf{Post-Test Score} }                                                                                                                                                                                   \\
Predictors                  & Estimates & CI                                 & Estimates & CI                                  & Estimates & CI                                  & Estimates & CI                                    \\ \hline
(Intercept)                 & $5.18^{\ast\ast\ast}$    & 4.48 – 5.89 &  $3.10^{\ast\ast\ast}$      & 1.85 – 4.36 &  $5.62^{\ast\ast\ast}$       & 5.03 – 6.21   & $5.60^{\ast\ast\ast}$      & 5.01 – 6.20  \\
Pre-Test Score              & $0.71^{\ast}$      & 0.04 – 1.37        &           &                                                  &              &                                     &              &                           \\
Attentive Self-Report Share &           &                                      & $0.06^{\ast\ast\ast}$      & 0.03 – 0.08 &           &                                        &           &                                         \\
Average KLD                 &           &                                       &           &                                       & 0.05      & -0.84 – 0.95                     &           &                                        \\
Average MM Similarity       &           &                                       &           &                                      &           &                                        & -0.22     & -1.13 – 0.69                     \\ \hline
Observations                & \multicolumn{2}{l}{84}                              & \multicolumn{2}{l}{84}                              & \multicolumn{2}{l}{84}                               & \multicolumn{2}{l}{84}                               \\
R2 / R2 adjusted            & \multicolumn{2}{l}{0.053 / 0.041}                   & \multicolumn{2}{l}{0.193 / 0.183}                   & \multicolumn{2}{l}{0.000 / -0.012}                   & \multicolumn{2}{l}{0.003 / -0.009}                   \\ \hline
\multicolumn{9}{r}{$^{\ast}$$ p\textless{}0.05    $$^{\ast\ast}$ p\textless{}0.01   $^{\ast\ast\ast}$ p\textless{}0.001}   
\end{tabular}
\end{adjustbox}
\end{table*}

\end{document}